\documentclass[twocolumn]{aastex62}
\usepackage[utf8]{inputenc}
\usepackage{savesym}
\savesymbol{tablenum}
\usepackage{siunitx}
\restoresymbol{SIX}{tablenum}
\usepackage[revtex,bfvec]{math-shortcuts}

\begin{document}

\title{Pulsar timing signatures of circumbinary asteroid belts}
\author{Ross J. Jennings}
\affiliation{Department of Astronomy, Cornell University, Ithaca, NY 14853, USA}
\author{James M. Cordes}
\affiliation{Department of Astronomy, Cornell University, Ithaca, NY 14853, USA}
\affiliation{Cornell Center for Astrophysics and Planetary Science, Cornell University, Ithaca, NY 14853, USA}
\author{Shami Chatterjee}
\affiliation{Department of Astronomy, Cornell University, Ithaca, NY 14853, USA}
\affiliation{Cornell Center for Astrophysics and Planetary Science, Cornell University, Ithaca, NY 14853, USA}

\begin{abstract}
The gravitational pull of a large number of asteroids perturbs a pulsar's motion to a degree that is detectable through precision timing of millisecond pulsars. The result is a low-frequency, correlated noise process, similar in form to the red timing noise known to affect canonical pulsars, or to the signal expected from a stochastic gravitational-wave background. Motivated by the observed fact that many millisecond pulsars are in binary systems, we describe the ways in which the presence of a binary companion to the pulsar would affect the signal produced by an asteroid belt. The primary effect of the companion is to destabilize the shortest-period orbits, cutting off the high-frequency component of the signal from the asteroid belt. We also discuss the implications of asteroid belts for gravitational-wave search efforts. Compared to the signal from a stochastic gravitational-wave background, asteroid belt noise has a similar frequency and amplitude, and is similarly independent of radio frequency, but is not correlated between different pulsars, which should allow the two kinds of signal to be distinguished.
\end{abstract}

\section{Introduction}

The high precision of pulsar timing makes it possible to detect orbital reflex motion created by small bodies orbiting pulsars, especially millisecond pulsars (MSPs). A striking demonstration of this is the discovery of pulsar planets. The smallest known pulsar planet, around PSR B1257+12, is less than twice as massive as the Moon \citep{wolszczan94,whk+00}, and a recent search for planets around MSPs \citep{brm+20} was able to rule out planets more massive than the Moon at periods as long as 100 days. Since such small bodies can be detected individually, it is worth considering the effects of even smaller bodies, such as asteroids, which can combine to produce a detectable signal.

The orbital reflex motion created by a body in a circular orbit around a pulsar produces a sinusoidal variation in pulse times of arrival (TOAs). For a body of mass $m$ in a circular orbit of radius $a$ around a pulsar of mass $M$, the amplitude, $\Delta\tau$, of this signal is
\begin{equation}\label{eqn:delta-tau}\begin{split}
    \Delta\tau&=\frac{ma}{Mc}\sin{i}\\
    &=\SI{50}{ns}\paren{\frac{m}{10^{-10}M}}\paren{\frac{a}{\SI{1}{au}}}\sin{i}.
\end{split}\end{equation}
Here $c$ is the speed of light and $i$ is the inclination of the orbital plane relative to the plane of the sky. If the ratio of the mass of the asteroid to the mass of the pulsar is $10^{-10}$ (around a typical pulsar, this corresponds to a mass similar to that of the large solar-system asteroid 4 Vesta) and it orbits at \SI{1}{au}, the signal amplitude can be as large as \SI{50}{ns}. This is large enough that a single asteroid could be detectable if it orbited one of the best-timed MSPs. The incoherent superposition of many such signals produced by an asteroid belt can be significantly stronger -- a belt with a total mass of only \num{5e-4}\,$M_\earth$ can produce a signal in excess of 250\,ns. For this reason, \citet{scm+13} suggested that an asteroid-belt signal could explain some of the low-frequency noise observed in TOAs from the bright MSP PSR B1937+21. Indeed, asteroid belts could be an important source of timing noise in MSPs in general, including in the large number of MSPs occurring in binary systems.

Asteroid belts around pulsars may be formed by supernova fallback material, or, in the case of MSPs, by material left over from the recycling process that spun the pulsar up. The latter is often thought to be the origin of the planets around PSR B1257+12. In addition to these, there are a number of other lines of evidence supporting the existence of debris disks or asteroid belts around at least some pulsars. The magnetars 4U~0142+61 \citep{wck06} and 1E~2259+586 \citep{kcw+09} exhibit infrared emission suggestive of debris disks. Furthermore, theoretical work related to the B1257+12 system \citep{mh01,bbr+06} indicates that asteroids larger than ~\SI{1}{km} in radius can survive for \SI{1}{Gyr} or more around an MSP.

Since many MSPs are found in binary systems with white dwarf companions, we are lead to consider additional effects that may arise in the binary context. Small objects such as asteroids orbiting in a binary star system have one of two types of stable orbits: satellite (S-type) orbits, circling one of the components; or planetary (P-type) orbits, encircling both components \citep{md99}. To maintain stability, S-type orbits must have much shorter periods (and smaller semi-major axes) than P-type orbits: there is an outermost stable S-type orbit and an innermost stable P-type orbit, which has a longer period. Combined with the scaling of TOA variations with orbital radius given by equation~(\ref{eqn:delta-tau}), this means that objects on P-type orbits will generally contribute more power to low-frequency TOA variations. In the remainder of this paper we will concern ourselves primarily with P-type asteroid belts.

\section{Spectral characteristics}\label{sec:spectral}

\begin{figure}
    \centering
    \includegraphics[width=\linewidth]{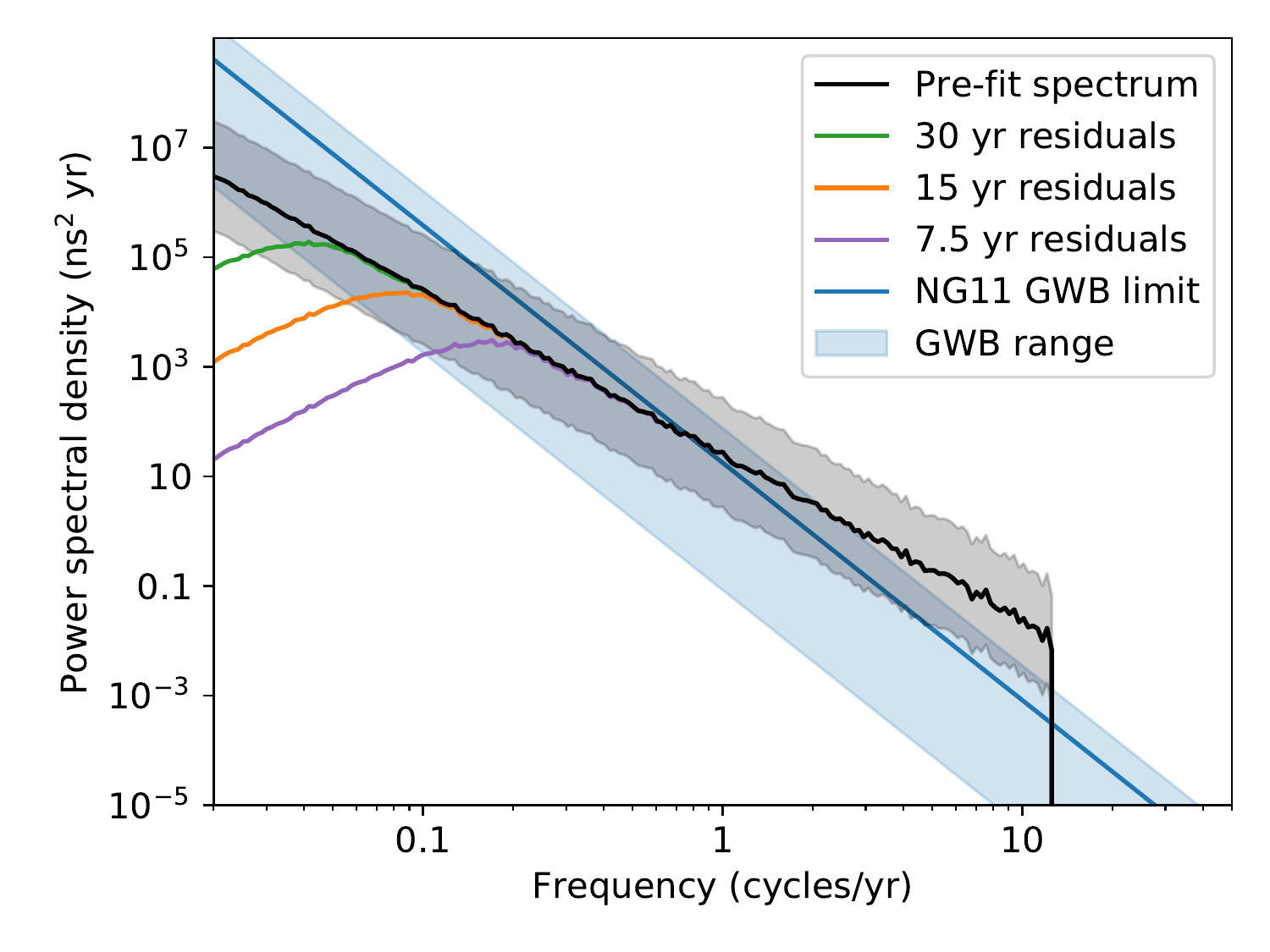}
    \caption{An example of the power spectrum of TOA perturbations  created by a P-type asteroid belt around a pulsar and its companion. The gray shaded region corresponds to an order of magnitude change, in either direction, in the asteroid belt mass. The residual spectrum is shown for 7.5, 15, and 30 year dataset lengths, demonstrating the effects of model subtraction (see section~\ref{sec:model-subtraction}). The spectrum expected from a GW background is shown for comparison, with the solid line corresponding to the upper limit on the strain amplitude, $A_{\mr{GWB}}=\num{1.45e-15}$, set by \citet{11yr-stochastic-background}, and the shaded region to $A_{\mr{GWB}}$ between \num{1e-16} and \num{3e-15}.}
    \label{fig:example-spectrum}
\end{figure}

A single asteroid of mass $m$ orbiting a pulsar of mass $M$ on a circular orbit of semimajor axis $a$ inclined at an angle $i$ from the plane of the sky produces a sinusoidal timing delay with amplitude given by equation~(\ref{eqn:delta-tau}).
The effects of additional asteroids on the timing residuals are additive as long as interactions between the asteroids can be neglected. When a large number of asteroids are present, they combine to produce a signal with an apparently continuous spectrum, as shown in Figure~\ref{fig:example-spectrum}.

The semimajor axis, $a$, of a particular asteroid is related to its orbital frequency, $f$, by
\begin{equation}\label{eqn:a}
    a=\paren{\frac{GM}{4\pi^2}}^{1/3}f^{-2/3}.
\end{equation}
This relationship is exact in the two-body problem with $M$ the sum of the masses. If one body is much less massive, as is the case for an asteroid orbiting a pulsar, it holds approximately with $M$ the mass of the larger body. In a binary system, it holds approximately for both S-type and P-type orbits with appropriate choice of $M$. For S-type orbits, the appropriate $M$ is the mass of the primary, and for P-type orbits, it is the total mass of the binary. The approximation becomes less accurate as the asteroid's orbit approaches the boundary of stability, but numerical evidence \citep{hw99,nagler05} suggests that orbits typically become unstable before deviations from equation~(\ref{eqn:a}) are significant. 

The spectrum of the TOA variations introduced by a large number of asteroids is related to the distribution of asteroid masses and orbital periods. Suppose that the number of asteroids with mass between $m$ and $m+\dd m$ and orbital frequency between $f$ and $f+\dd f$ is
\begin{equation}
    dN=n\of{m, f}\dd m\dd f.
\end{equation}
The number for which the amplitude is between $\tau$ and $\tau+\dd\tau$ and the orbital frequency is between $f$ and $f+\dd f$ is therefore
\begin{equation}
    \dd N = n_*\of{\tau, f}\dd\tau\dd f,
\end{equation}
where
\begin{equation}\label{eqn:nstar}
    n_*\of{\tau, f}=K f^{2/3}\,n\of{K f^{2/3}\tau, f},
\end{equation}
and
\begin{equation}
    K=\paren{\frac{4\pi^2M^2}{G}}^{1/3}\frac{c}{\sin{i}}.
\end{equation}
The power spectral density of the TOA signal, $S(f)$, is calculated by summing the contributions from each asteroid. A single asteroid contributes power $\tau^2/4$ to the power spectrum at frequency $f$, so the power spectral density (defined to contain only positive frequencies) is given by
\begin{equation}
    S\of{f}=\frac14\int_0^\infty \tau^2 n_*\of{\tau, f}\dd\tau.
\end{equation}
If the distribution of asteroids has a power-law form in mass and frequency, with upper (lower) cutoffs $m_+$ ($m_-$) in mass and $f_+$ ($f_-$) in frequency, $n\of{m,f}$ is given by
\begin{equation}\label{eqn:powlawn}
    n\of{m, f} = \frac{N\alpha\beta\, m^{\alpha-1}f^{\beta-1}}{\paren{m_+^\alpha-m_-^\alpha}\paren[big]{f_+^\beta-f_-^\beta}},
\end{equation}
where $m_- < m < m_+$ and $f_- < f < f_+$, and $0$ elsewhere. The average power spectral density at frequency $f$ is therefore
\begin{equation}\label{eqn:s-of-f}
    S\of{f}=\paren{\frac{G}{4\pi^2M^2}}^{2/3}\frac{M_{\mr{belt}}m_{\mr{eff}}}{4c^2}\frac{\beta\,f^{\beta-7/3}}{\paren[big]{f_+^\beta-f_-^\beta}}\sin^2{i},
\end{equation}
for $f_-<f<f_+$, and $0$ otherwise. Here $M_{\mr{belt}}$ is the total mass of the belt, and $m_{\mr{eff}}$ is effective mass of a single asteroid, given by
\begin{equation}\label{eqn:meff}
    m_{\mr{eff}}\of\alpha=\frac{\bracket{m^2}}{\bracket{m}}=\frac{\paren{\alpha+1}\paren{m_+^{\alpha+2}-m_-^{\alpha+2}}}{\paren{\alpha+2}\paren{m_+^{\alpha+1}-m_-^{\alpha+1}}},
\end{equation}
which always lies between the minimum and maximum masses, $m_-$ and $m_+$. When $m_+\gg m_-$, $m_{\mr{eff}}$ is dominated by $m_+$ for $\alpha>-1$ and by $m_-$ for $\alpha<-2$. For $-2<\alpha<-1$, it behaves as a weighted geometric mean of the two. 

An example spectrum is shown in Figure~\ref{fig:example-spectrum}. It is an ensemble average calculated from $10^4$ realizations of an asteroid belt containing $10^4$ asteroids, distributed in mass and semimajor axis according to equation~(\ref{eqn:powlawn}). The overall shape of the spectrum is described by equation~(\ref{eqn:s-of-f}). The high-frequency cutoff corresponds to the period of the innermost stable P-type orbit in the PSR J1614$-$2230 system, determined using the formula of \citet{hw99} (equation \ref{eqn:ap}). The total mass of the belt is chosen to match that of the solar system's asteroid belt. The spectral indices in mass and frequency are $\alpha=-5/6$ and $\beta=-2/3$, respectively, and the total number of asteroids, $N$, is $10^4$. The upper and lower cutoffs in frequency are $f_-=\SI{0.0173}{yr^{-1}}$ and $f_+=\SI{12.5}{yr^{-1}}$, and the cutoffs in mass are $m_-=\num{8.4e-15}\,M_\sun$ and $m_+=\num{1.0e-10}\,M_\sun$. These are the same as the parameter values used for PSR J1614$-$2230 in Figure \ref{fig:spectra}. More information on why these parameter values were selected can be found in section~\ref{sec:results}.

Notably, the shape of the spectrum described by equation~(\ref{eqn:s-of-f}) depends only on the distribution of asteroids in orbital frequency. In particular, it has the form $S\of{f}\propto f^{-\gamma}$, where $\gamma=\frac73-\beta$. The mass distribution enters only through $m_{\mr{eff}\of{\alpha}}$, which scales the spectrum linearly. The effect of $\beta$ on the shape is shown in Figure~\ref{fig:slope-comparison}. In addition to the value used in Figure~\ref{fig:example-spectrum} ($\beta=-2/3$), spectra corresponding to two other values of $\beta$ are shown: $\beta=2/3$, which corresponds to a uniform disk surface density, and $\beta=-2$, which produces the same spectral slope, $\gamma=13/3$, expected from the stochastic gravitational-wave (GW) background.

The total variance that asteroids contribute to the TOAs is
\begin{equation}
    \sigma_\tau^2 = \int_{-\infty}^\infty S\of{f}\dd f = 2\int_0^\infty S\of{f}\dd f.
\end{equation}
For a power-law distribution (equation~\ref{eqn:powlawn}), this is
\begin{equation}\label{eqn:sigma-tau}
    \sigma_\tau^2=\paren{\frac{G}{4\pi^2M^2}}^{2/3}\frac{M_{\mr{belt}}m_{\mr{eff}}}{2c^2}\frac{\beta\paren[big]{f_+^{\beta-4/3}-f_-^{\beta-4/3}}}{\paren{\beta-\frac43}\paren[big]{f_+^\beta-f_-^\beta}}\sin^2{i}.
\end{equation}
Fixing the binary mass and asteroid frequency distribution to the values used in Figure~\ref{fig:example-spectrum}, this becomes
\begin{equation}\label{eqn:sigma-tau-fiducial}
    \sigma_\tau=\SI{170}{ns}\sqrt{\frac{N}{10^4}}\paren{\frac{m_{\mr{rms}}\sin{i}}{10^{-12}\,M_\sun}}.
\end{equation}
For the asteroid mass distribution used in Figure~\ref{fig:example-spectrum}, $m_{\mr{rms}}=\num{1.7e-12}\,M_\sun$, so the standard deviation of the TOA perturbations in the case shown there is approximately \SI{290}{ns}.

\subsection{Effects of model subtraction}\label{sec:model-subtraction}

\begin{figure}
    \centering
    \includegraphics[width=\linewidth]{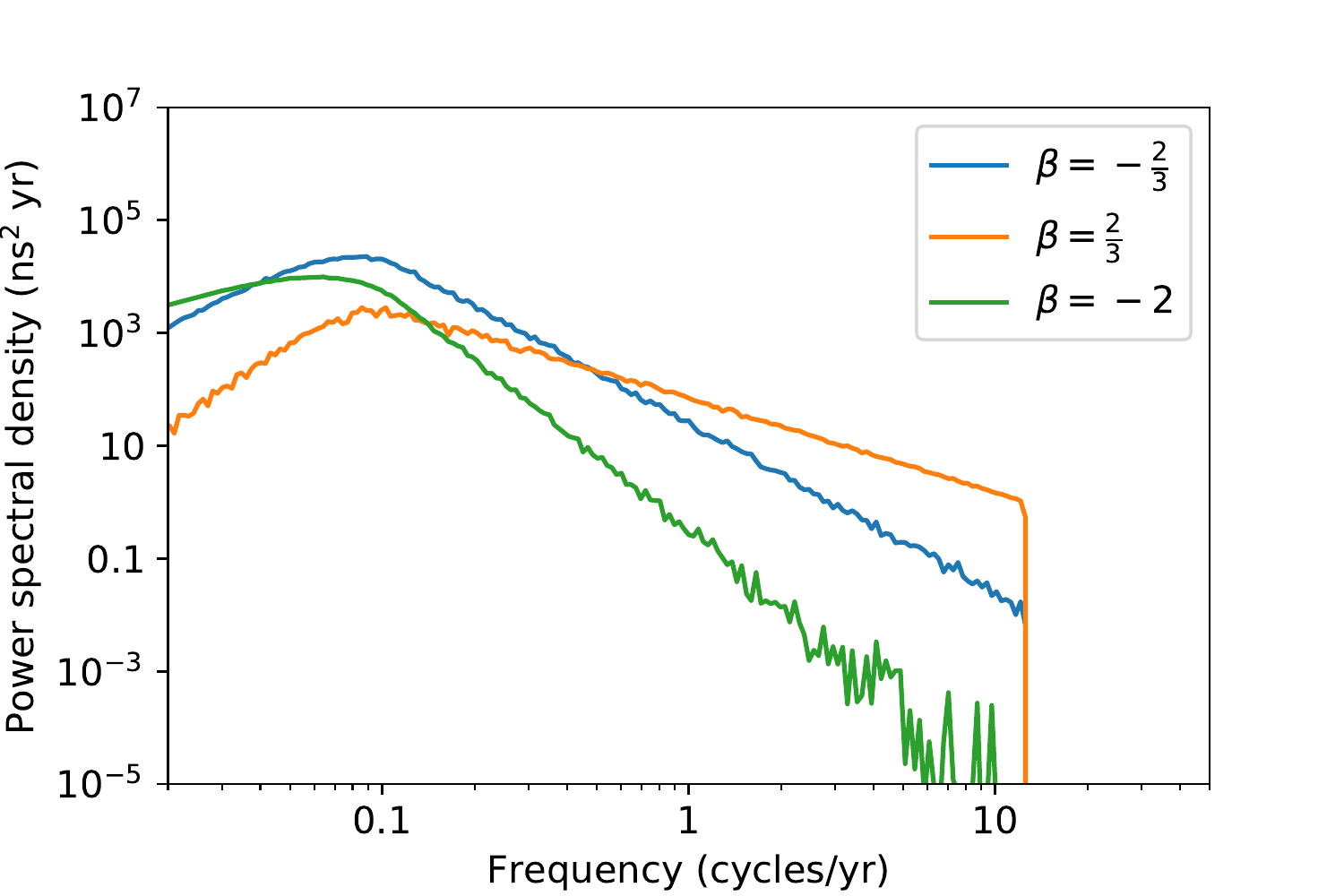}
    \caption{The effect of the power-law index, $\beta$, of the orbital frequency distribution of asteroids on the TOA residual spectrum. The $\beta$ values shown correspond to a uniform distribution in semimajor axis ($\beta=-2/3$, $\gamma=3$), a uniform disk surface density ($\beta=2/3$, $\gamma=5/3$), and a spectral slope equivalent to that expected from the stochastic gravitational-wave background ($\beta=-2$, $\gamma=13/3$). All other parameters used are the same as those in Figure~\ref{fig:example-spectrum}. The transmission function corresponding to a dataset length of 15 years has been applied in each case.}
    \label{fig:slope-comparison}
\end{figure}


Any high-precision pulsar timing analysis requires fitting a timing model describing the pulsar's position, spindown, parallax, proper motion, and, if the pulsar has a binary companion, its orbit. An additional signal not present in the timing model will often be partially degenerate with the timing model parameters, especially if the signal is spread over a wide range of frequencies. The spindown component, in particular, amounts to a polynomial trend, absorbing power at the lowest frequencies -- those with periods comparable to, or longer than, the length of the data set. This sets a lower limit on the frequency of observable signals.

The amount of power removed by subtracting the timing model is quantified by the transmission function, $\mc{T}\of{f}$, the factor by which the power at frequency $f$ is reduced \citep{bnr84,mcc13,hrs19}. For most pulsars, the spindown is described by a model in which the delay is a quadratic function of time. The transmission function for such a model behaves as
\begin{equation}
    \mc{T}\of{f}\sim\paren{\frac{f}{f_c}}^6,\quad f\lesssim f_c,
\end{equation}
when $f$ is significantly less than the critical frequency
\begin{equation}
    f_c=\frac{1575^{1/6}}{\pi T_{\mr{obs}}}\approx\frac{1.086}{T_{\mr{obs}}},
\end{equation}
where $T_{\mr{obs}}$ is the length of the data set. This causes the observed spectrum of the TOA residuals created by an asteroid belt to roll off at low frequencies, as illustrated in Figure~\ref{fig:example-spectrum} for various data set lengths. The same effect is applied in Figure~\ref{fig:slope-comparison}. For a timing model including position and astrometric parameters, the transmission function would also include a dip around $f=\SI{1}{yr^{-1}}$, but this has a much less significant effect on the total variance, so we have not included it in Figures \ref{fig:example-spectrum} and \ref{fig:slope-comparison}.

\section{Results}\label{sec:results}


\begin{deluxetable}{l r r r r r}
\tablecaption{Stability boundaries for some MSP binaries}
\tablehead{PSR & $P_{B}$ (days) & \multicolumn{2}{c}{$P_{S}$ (days)} & \multicolumn{2}{c}{$P_{P}$ (days)}\\
\cline{3-4}
\cline{5-6}
& & HW99 & Q+20 & HW99 & Q+18}
\startdata
J0437$-$4715 & $5.741$ & $1.65$ & $1.67$ & $16.7$ & $18.2$ \\
J1614$-$2230 & $8.687$ & $2.34$ & $2.03$ & $29.0$ & $28.3$ \\
J1713+0747 & $67.825$ & $18.7$ & $15.8$ & $217$ & $220$ \\
J1741+1351 & $16.335$ & $4.56$ & $4.41$ & $50.9$ & $52.6$ \\
B1855+09 & $12.327$ & $3.47$ & $3.46$ & $37.7$ & $40.5$ \\
J1903+0327 & $95.174$ & $7.40$ & $5.58$ & $664$ & $634$ \\
J1909$-$3744 & $1.533$ & $0.441$ & $0.451$ & $4.44$ & $4.87$ \\
J1918$-$0642 & $10.913$ & $3.07$ & $3.07$ & $33.4$ & $34.9$ \\
J2043+1711 & $1.482$ & $0.431$ & $0.440$ & $4.2$ & $4.67$ \\
\enddata
\tablecomments{\label{tab:example-systems} $P_S$ is the period of the outermost stable circular orbit around the pulsar, and $P_P$ is the period of the innermost stable circular orbit around the binary. The orbital period, $P_B$, of each pulsar is included for reference. The estimates labeled HW99 are based on the results of \citet{hw99} and those labeled Q+18 and Q+20 are based on those of \citet{qsk+18} and \citet{qlk+20}, respectively. For further details, see Appendix~\ref{sec:appendix}.}
\end{deluxetable}

\begin{figure*}
    \centering
    \includegraphics[width=\textwidth]{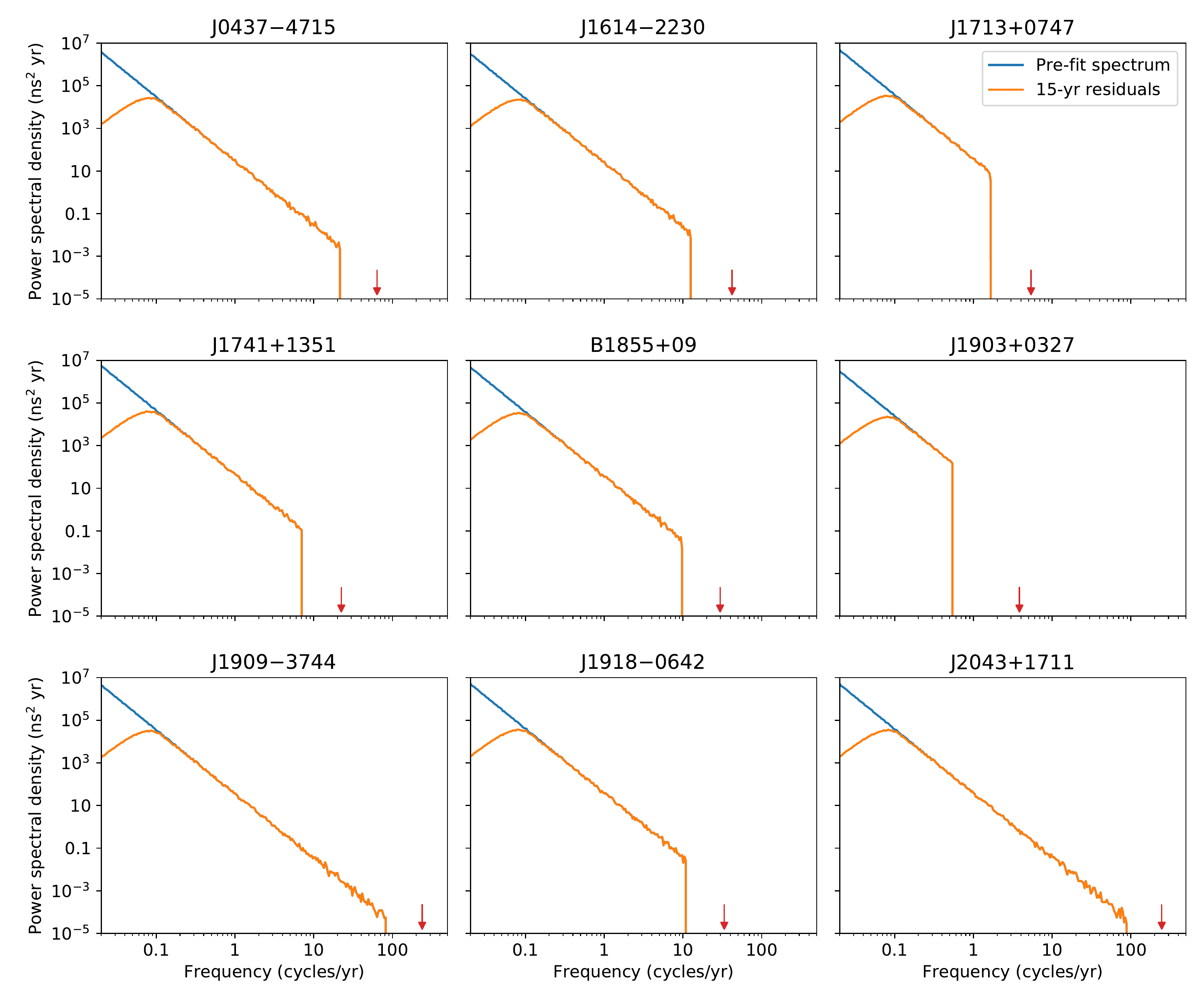}
    \caption{Simulated spectra of TOA perturbations for model P-type asteroid belts around each of the systems in Table~\ref{tab:example-systems}. In each case, the spectrum is shown both before and after the subtraction of a quadratic spindown model for a nominal dataset length of 15 years, and the small red arrow indicates the pulsar's orbital frequency. As in Figure~\ref{fig:example-spectrum}, the total mass of the model asteroid belt in each case is chosen to match that of the solar system's asteroid belt, and the distribution of asteroids in mass and orbital frequency has a power-law form (cf. equation~\ref{eqn:powlawn}), with indices $\alpha=-5/6$ in mass and $\beta=-2/3$ in frequency. More information on how parameters were selected can be found in section~\ref{sec:results}.}
    \label{fig:spectra}
\end{figure*}

The boundaries of stability for several MSP binary systems are given in Table~\ref{tab:example-systems}. For each pulsar system, two estimates of the stability boundary are given, one using the interpolating polynomials given by \citet{hw99} (equations~\ref{eqn:as} and~\ref{eqn:ap}), and the other based on the results of \citet{qsk+18} and \citet{qlk+20}. We give further context on these estimates in Appendix~\ref{sec:appendix}. The systems in the table were selected because measurements of the pulsar and companion masses are available in the literature. All but one have almost perfectly circular orbits, a common characteristic of pulsar-white dwarf binaries. The exception, PSR J1903+0327, has a main-sequence companion on an eccentric orbit ($e=0.437$). Simulated spectra of (P-type) asteroid belts around each system are shown in Figure~\ref{fig:spectra}, demonstrating the relationship between the binary orbital frequency and the frequency of the innermost stable P-type orbit. 

While all the systems in Table~\ref{tab:example-systems} have orbital periods between 1 and 100 days, the orbital periods of known pulsar--white dwarf binary systems range from 2.46 hours in the case of PSR J0348+0432 \citep{afw+13} to 3.37 years in the case of PSR B0820+02 \citep{hlk+04}. Even more extreme orbital periods are found in 47 Tuc R \citep{clf+00}, which has an ultra-light companion in a 96-minute orbit, and PSR J2032+4127 \citep{hnl+17}, which has a main-sequence companion in a highly eccentric 46-year orbit. The inner boundary of stability has a similar range. Using the minimum companion mass in each case, we find that it corresponds to a period of 6 hours for PSR J0348+0432 and 10 years for PSR B0820+02. For the longest-period objects, S-type asteroid belts may be more relevant than their P-type counterparts, but we will not consider them in detail here.

Each of the spectra seen in Figure~\ref{fig:spectra} is an average of $10^4$ realizations of an asteroid belt containing $10^4$ asteroids with masses and orbital frequencies drawn at random from power-law distributions of the form given by equation~(\ref{eqn:powlawn}). In each case the belt has a total mass of \num{2e-9} solar masses (\num{6.7e-4} earth masses), comparable to that of the solar system's asteroid belt. The maximum orbital frequency, $f_+$, is chosen to match the frequency of the innermost stable P-type orbit in the particular binary system, as given in Table~\ref{tab:example-systems}. The minimum orbital frequency, $f_-$, corresponds in each case to a semimajor axis of \SI{20}{au}. This was chosen, somewhat arbitrarily, to make the longest orbital period longer than the 15-year nominal dataset length, since there are few constraints on how far a P-type asteroid belt extends away from the binary system. Increasing the minimum orbital frequency narrows the range of frequencies over which the asteroid signals are distributed, and, if the total mass of the belt is held fixed, also increases the density of asteroids in orbital frequency space. Both of these effects tend to make the asteroid belt signal more recognizable.

As demonstrated in Figure~\ref{fig:slope-comparison}, the shape of the spectrum is primarily determined by the orbital frequency distribution. In every case shown in Figure~\ref{fig:spectra}, the power-law index of this distribution is $\beta=-2/3$, which represents a uniform distribution in semimajor axis. This is the same as what is used in Figure \ref{fig:example-spectrum}.

The value $\alpha=-5/6$ is used for the power-law index of the mass distribution throughout Figure~\ref{fig:spectra}. This was chosen to match the value predicted by \citet{dohnanyi69} for material in collisional equilibrium. As equations~(\ref{eqn:s-of-f}) and~(\ref{eqn:meff}) demonstrate, the mass distribution affects the ensemble average spectrum only by determining the relationship between the mass cutoffs $m_+$ and $m_-$, the RMS mass $m_{\mr{rms}}$, and the total mass. However, in single realizations, the mass distribution is important because it determines how common asteroids with masses much larger than $m_{\mr{rms}}$ are. This has important implications for detecting individual asteroids (see section~\ref{sec:detection}).



\section{Discussion}

The results shown so far are based on a model asteroid belt whose total mass is equal to that of the solar system's asteroid belt. As shown in equation~\ref{eqn:s-of-f}, the amplitude of the TOA signal produced by the belt scales linearly with both the total mass of the belt and the mass of a typical asteroid.

\begin{figure*}
    \centering
    \includegraphics[width=\textwidth]{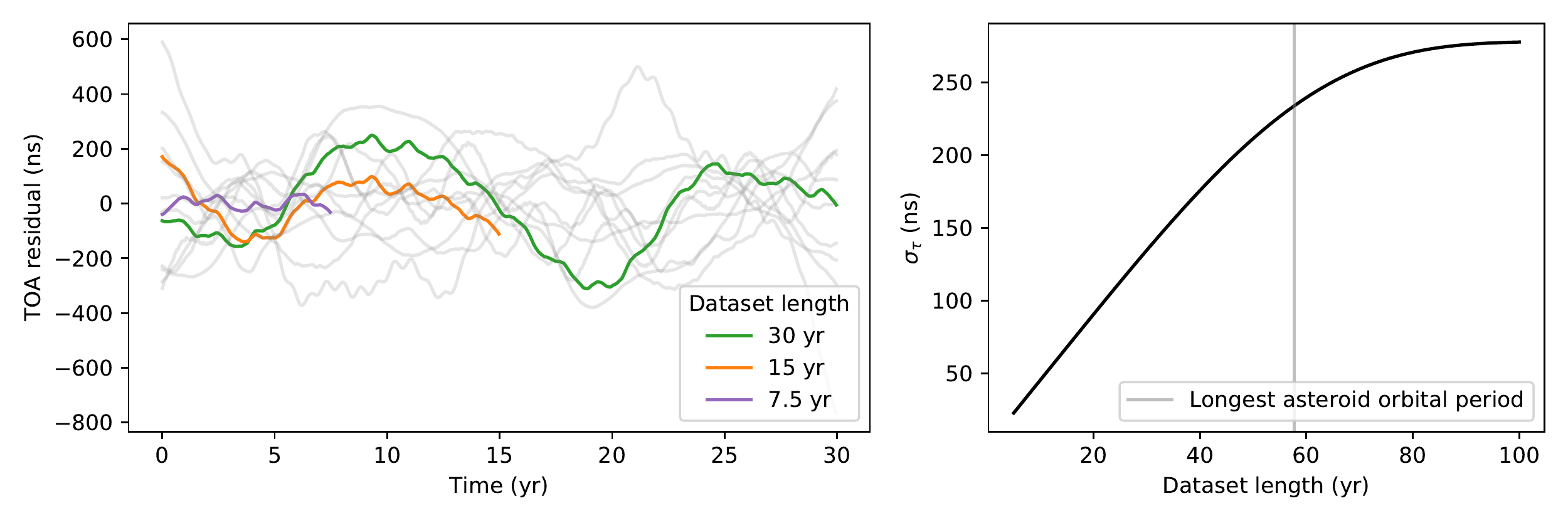}
    \caption{Example time series of TOA residuals created by an asteroid belt (after timing model subtraction), illustrating how the variance of the residuals increases with dataset length. The light gray lines correspond to different realizations with the same parameters used for Figure~\ref{fig:example-spectrum}, with a quadratic trend removed. For one particular realization, highlighted in color, the first 7.5 and 15 years of the time series have been isolated, and the fitting and subtraction of the trend performed separately. This is quantified in the right panel, which shows how the standard deviation, $\sigma_{\tau}$, of the residuals behaves as a function of the dataset length in the ensemble average. While $\sigma_{\tau}$ initially grows linearly with time, it eventually stops growing and stabilizes at approximately \SI{290}{ns}, the value predicted by equation~(\ref{eqn:sigma-tau}). This happens after the dataset length exceeds the longest asteroid orbital period.}
    \label{fig:time-series}
\end{figure*}

\subsection{Comparison with other low-frequency timing signals}

An asteroid belt is hardly the only phenomenon capable of producing red noise in pulsar timing residuals. In fact, red noise of a much larger amplitude than that considered here is common in canonical pulsars (those with periods of order one second and period derivatives of order $10^{-15}$, as opposed to MSPs, which have periods of a few milliseconds and period derivatives of order $10^{-20}$), and is generally understood to arise from stochastic variations in the pulsar's rotation rate, caused by some combination of magnetospheric torques \citep{klo+06,lhk+10} and instabilities arising from differential rotation between the neutron star's surface and its superfluid interior \citep{jones90}. It is precisely the fact that MSPs are relatively free of this kind of timing noise that, combined with their shorter periods, makes them ideal for high-precision timing applications, including GW searches.

The amplitude of red timing noise in canonical pulsars has been observed to scale with the period derivative, $\dot{P}$~\citep{ch80,antt94}, making it unsurprising that the effect should be smaller in MSPs. Indeed, \citet{sc10} developed a scaling relation that is consistent with the lower levels of red noise observed in MSPs as well as the higher levels observed in canonical pulsars, indicating that they may have the same origin. \citet{lcc+17} followed this up with a fit including more MSPs, again finding similar behavior between canonical pulsars and MSPs. However, it remains possible that at least some of the observed red noise in MSPs may have another origin.

Perturbations caused by encounters with nearby stars, or with smaller objects such as interstellar planets or asteroids, could also give rise to low-frequency pulsar timing signals. However, encounters with stars typically take place too slowly for changes in the acceleration of the pulsar to be detectable \citep{phinney93}, and detectable encounters with smaller objects are rare events with a distinctive signature \citep{jcc20}. The effects of both kinds of encounters can therefore be distinguished from the red noise produced by an asteroid belt.

\subsection{Stationarity}\label{sec:stationarity}

The TOA perturbations generated by an asteroid belt can be compared to those produced by random walks in pulse phase, frequency, or frequency derivative, all of which are statistically non-stationary. However, the signal from an asteroid belt is in principle stationary when measured over a long enough span of time --- there is always a longest-period asteroid. If the asteroids have a power-law distribution in orbital frequency (equation~\ref{eqn:powlawn}), so that the spectrum of the TOA perturbations also has a power law form, the variance of the residuals will grow with dataset length, but only up to a certain point,  determined by the low-frequency (long-period) cutoff of the distribution. This can be seen in Figure~\ref{fig:time-series}. In the case shown there, where asteroids are distributed uniformly in semimajor axis, the variance initially grows linearly with time, but other power-law indices are possible depending on the asteroid distribution. As discussed in section~\ref{sec:spectral}, if the  power-law index of the orbital frequency distribution is $\beta$, the spectrum will have the form $S\of{f}\propto f^{-\gamma}$, where $\gamma=\frac73-\beta$ (equation~\ref{eqn:s-of-f}). In this case, the standard deviation of the TOA residuals will increase with time as $\sigma_{\mr{TOA}}\propto T^{\delta}$, with $\delta=\frac12(\gamma-1)=\frac23-\frac12\beta$. A uniform distribution in semimajor axis gives $\gamma=3$, while a uniform surface density gives $\gamma=5/3$. For comparison, random walks in pulse phase, frequency, and frequency derivative correspond to $\gamma=2$, $4$, and $6$, respectively, and the expected spectrum of the stochastic gravitational-wave background corresponds to $\gamma=13/3$.

\subsection{Survival of asteroids}

Several factors other than the presence of a binary companion affect the stability of asteroid belts around pulsars. Many of these are especially relevant for millisecond pulsars, which are generally billions of years old. The seasonal Yarkovsky effect \citep[e.g.][]{rubincam98}, a thermal drag mechanism acting on objects whose spin axes are tilted relative to their orbits, can cause small asteroids to migrate inward until they are evaporated. However, the migration timescale is proportional to asteroid radius, and \SI{5}{km} asteroids can survive for at least \SI{250}{Myr} at \SI{1}{au} \citep{cs08,scm+13}.

An asteroid will be evaporated if it is heated enough for its equilibrium temperature to exceed the melting point of the materials that compose it. Asteroids in the vicinity of a pulsar are heated by a number of mechanisms in addition to thermal emission from the pulsar's surface, including particle and X-ray emission driven by the pulsar's spindown and Ohmic dissipation driven by currents between the asteroid and the pulsar's magnetosphere. A detailed analysis of these effects was carried out by \citet{cs08}, who reached the conclusion that asteroids do not begin to evaporate until they are within approximately $10^{10}\,\si{cm}$ (\SI{6e-4}{au}) of the pulsar, which corresponds to a seven-minute orbital period, or a frequency of \num{7e5} cycles per year. It follows that asteroids on stable P-type orbits in pulsar binary systems, even those with the shortest orbital periods, are almost certainly safe from evaporation.


For asteroids that are massive enough or close enough together, their mutual gravitational interactions can also be a destabilizing influence. Examining this issue, \citet{ht10} conclude that the timescale for destabilization in ``cold'' disks increases exponentially with the separation between the asteroids, with a distance of around 10 times the Hill radius sufficing for a lifetime of \SI{100}{Myr}.

In globular clusters, stellar encounters can limit the outer radius of an asteroid belt. An asteroid's orbit is likely to be disrupted if another star approaches the pulsar to within a few times the orbit's radius. In a cluster with stellar density $n_*$ and velocity dispersion $\sigma$, the rate of encounters with periapsis distance less than $r$ is given by
\begin{equation}
    \Gamma_*\of{r}=\pi r^2\sigma n_* \paren{1+\frac{2GM}{\sigma^2r}}
\end{equation}
\citep[cf.][]{vh87}, where the factor in parentheses accounts for enhancement of the encounter cross-section due to gravitational focusing, and $M$ is the total mass of the interacting bodies. For a typical globular cluster with $n_*=10^4\,\si{pc^{-3}}$ and $\sigma=\SI{10}{km.s^{-1}}$, encounters with $r<\SI{10}{au}$ happen approximately once every \SI{250}{Myr}, assuming typical masses of \num{1.0} and \num{1.4}\,$M_\sun$ for the star and pulsar, respectively. In the less dense environment of the galactic disk, where the stellar density is some four orders of magnitude smaller, encounters are much rarer, and even at 100 AU are unlikely to happen in the lifetime of the pulsar.

\subsection{Distinguishing individual asteroids}\label{sec:detection}

One way to test the hypothesis that observed red noise in TOAs from a pulsar is caused by an asteroid belt is to separate out the signals caused by individual asteroids. This is possible only if the dataset is long enough to acquire sufficient resolving power in the frequency domain. If there are only a handful of asteroids, it may be possible to observe for long enough to resolve each individual asteroid in frequency space, but for denser asteroid belts this becomes impractical. The average number, $\Delta N$, of asteroids in a single frequency bin is related to the density,
\begin{equation}\label{eqn:dn-df}
    \der{N}{f}=\frac{N\beta f^{\beta-1}}{f_+^\beta-f_-^\beta},
\end{equation}
of asteroids in frequency space by
\begin{equation}
    \Delta N=\der{N}{f}\Delta f,
\end{equation}
where $\Delta f$ is the width of the frequency bin. Asteroids will be individually resolvable if $\Delta N\lesssim 1$. Because the minimum $\Delta f$ achievable with a dataset of length $T$ is approximately $1/T$, this means that asteroids will be individually resolvable in frequency only if
\begin{equation}
    T\gtrsim\der{N}{f}.
\end{equation}
Using the parameters from section~\ref{sec:spectral}, this becomes
\begin{equation}\label{eqn:dn-df-scaling}
    T\gtrsim\SI{450}{yr}\paren{\frac{N}{10^4}}\paren{\frac{f}{\SI{1}{yr^{-1}}}}^{-5/3}.
\end{equation}
This suggests that realistic observations will usually be in the opposite regime, in which $\Delta N\gg 1$. In such cases, the central limit theorem may be applied to individual frequency bins, so the probability distribution of the complex signal amplitude in each bin will be Gaussian, and the probability distribution of the power will be exponential, with mean $S\of{f}\Delta f$. Since the standard deviation of an exponential distribution is equal to its mean, the standard deviation of the power in the frequency bin centered at $f$ will also be $S\of{f}\Delta f$.

An additional sinusoidal signal may be considered distinguishable from the asteroid belt signal if the probability of it arising by chance in the asteroid belt model is sufficiently small. For a signal of amplitude $\Delta\tau$, the ratio of the power in the signal to power in asteroid belt noise is
\begin{equation}\label{eqn:snr}
    \mr{S}/\mr{N}=\frac{(\Delta\tau)^2}{4S\of{f}\Delta f},
\end{equation}
where $\Delta f$ is the frequency resolution. The signal may be said to be detected if the signal-to-noise ratio exceeds a threshold value, $s$. Under the null hypothesis $\Delta\tau=0$, the probability that the threshold will be exceeded is
\begin{equation}
    p=\exp\of{-s}.
\end{equation}
If $N_f$ frequency bins are examined, the probability that at least one will exceed the threshold value by chance is
\begin{equation}
    P=1-(1-p)^{N_f}\approx {N_f}p.
\end{equation}
It follows that, to achieve a false positive probability of $P$, the threshold should be set to
\begin{equation}
    s=\ln\of{\frac{N_f}{P}}.
\end{equation}
The signal is then detectable if
\begin{equation}
    \frac{(\Delta\tau)^2}{4S\of{f}\Delta f}\gtrsim\ln\of{\frac{N_f}{P}}.
\end{equation}
In the case that the signal is created by another asteroid, so that $\Delta\tau$ is given by equation~(\ref{eqn:delta-tau}), and the asteroids in the belt have power law distributions in mass and orbital frequency (equation~\ref{eqn:powlawn}), so that $S\of{f}$ is given by equation~(\ref{eqn:s-of-f}), the signal-to-noise ratio (equation~\ref{eqn:snr}) becomes
\begin{equation}
    \mr{S}/\mr{N}=\frac{Nm^2}{M_{\mr{belt}}m_{\mr{eff}}} T\paren{\der{N}{f}}^{-1}.
\end{equation}
Here $m_{\mr{eff}}$ is given by equation~(\ref{eqn:meff}), and $\dd{N}/\dd{f}$ is the density of asteroids in orbital frequency (equation~\ref{eqn:dn-df}).
An additional asteroid of mass $m$ is therefore distinguishable from the bulk of the belt if
\begin{equation}
      m\gtrsim\sqrt{\frac{\beta f^{\beta-1}}{\paren[big]{f_+^\beta-f_-^\beta}T}\ln\of{\frac{N_f}{P}}}\sqrt{M_{\mr{belt}} m_{\mr{eff}}}.
\end{equation}

Taking $N_f=100$ and $P=0.05$ (so that the signal-to-noise threshold is $s=7.6$), and using the parameters from section~\ref{sec:spectral}, this becomes $m \gtrsim m_{\mr{thres}}$, where
\begin{equation}
    m_{\mr{thres}}= \num{6.7e-11}\,M_\sun\paren{\frac{f}{\SI{1}{yr^{-1}}}}^{-5/6}\paren{\frac{T}{\SI{15}{yr}}}^{-1/2}.
\end{equation}
The expected number of asteroids with masses greater than the threshold, $m_{\mr{thres}}$, is
\begin{equation}
   N_{>}=\frac{N\paren[big]{m_+^\alpha-m_{\mr{thres}}^\alpha}}{m_+^\alpha-m_-^\alpha}.
\end{equation}
For the parameters in section~\ref{sec:spectral}, this is
$N_>=11$. In other words, in an asteroid belt like the one shown in Figure~\ref{fig:example-spectrum}, approximately 11 of the $10^4$ asteroids would be detectable individually in a 15-year dataset, assuming that the uncertainties in the TOA measurements are negligible compared to the signal from the rest of the asteroid belt.

\subsection{Implications for gravitaional-wave searches}

In a number of ways, the pulsar timing signal produced by an asteroid belt closely resembles the signal expected from gravitational wave (GW) sources. Like the stochastic GW background, asteroid belts around pulsars should produce correlated noise in TOAs with frequencies of order \SI{1}{yr^{-1}} and amplitudes of tens of nanoseconds; and, like continuous wave sources, individual large asteroids should produce approximately sinusoidal TOA perturbations. Both GW and asteroid-belt signals can be distinguished from the effects of dispersion and scattering produced by the interstellar medium in that they are achromatic, i.e., they do not depend on radio frequency. This means that unlike, for example, TOA variations caused by changes in dispersion measure (DM), variations caused by the presence of an asteroid belt cannot be measured and corrected for by comparing signals at different radio frequencies.

However, there is at least one important way in which asteroid-belt and GW signals differ. In particular, the Earth-term component of any GW signal should be correlated across different pulsars, with a characteristic spatial pattern, originally described by \citet{hd83}, that arises from the quadrupolar nature of GWs. Asteroid belts, on the other hand, belong to particular pulsars; even if all pulsars had asteroid belts with identical statistical properties, there would be no reason to expect the signals they produced to be correlated, since the masses, phases, and orbital frequencies of individual asteroids would differ from one pulsar to the next.

\section{Summary and Conclusions}

Asteroid belts in pulsar systems produce achromatic, low-frequency timing noise, which arises from the orbital reflex motion of the pulsar. The presence of a binary companion in a pulsar system has the effect of destabilizing orbits close to the pulsar. This provides a natural upper cutoff for the orbital frequencies of any asteroids, but does not exclude the possibility of an asteroid belt entirely, since sufficiently distant orbits remain stable. 

The hypothesis that observed timing noise in one or more MSPs is produced by an asteroid belt may be tested by looking for evidence of stationarity, which should be present only if the frequency corresponding to the outer edge of the belt is observable; or by trying to isolate the signal from an individual large asteroid. Completely resolving asteroids in orbital frequency is possible only for very sparse asteroid belts or very long datasets, but it may be possible to detect individual large asteroids well before this point. Nevertheless, in the near term it is likely to remain challenging to determine whether particular instances of pulsar timing noise are caused by asteroid belts.

Asteroid belts are potentially important as noise sources in searches for low-frequency gravitational waves. The TOA signal produced by an asteroid belt is similar to that expected from a GW background in its frequency, amplitude, and achromatic nature, and may have a similar power-law spectrum. However, it differs in that it is not expected to be correlated between different pulsars. This makes it particularly important for GW searches to consider the correlation between signals in different pulsars.

\acknowledgments

The authors are members of the NANOGrav Physics Frontiers Center, which receives support from the National Science Foundation (NSF) under award number 1430284. This paper made use of data from the ATNF pulsar catalogue\footnote{https://www.atnf.csiro.au/research/pulsar/psrcat/} \citep{mht+05}. We thank D. Lai for useful comments regarding orbital stability.

\appendix

\section{Stable orbits in a binary system}\label{sec:appendix}

In any binary system, there is an approximate maximum radius for stable S-type orbits and an approximate minimum radius for stable P-type orbits. The boundary of the stable region is actually irregular and fractal in nature, with ``teeth'' corresponding to resonant orbits~\citep{nagler05,shevchenko15}, but this approximation will suffice for our purposes.


\citet{hw99} conducted a numerical investigation of the stability of circular S- and P-type orbits for various combinations of the eccentricity $e$ and mass ratio $\mu$ of the central binary. They found that S-type orbits with semi-major axes less than a critical value $a_S$, and P-type orbits with semi-major axes greater than a critical value $a_P$, were stable for the duration of their simulations ($10^4$ orbital periods). They give the maximum semi-major axis for stable S-type orbits as a quadratic polynomial in $e$ and $\mu$:
\begin{equation}\label{eqn:as}\begin{split}
    a_S &= [(0.464\pm0.006)+(-0.380\pm0.010)\mu\\
    &\quad+(-0.631\pm0.034)e+(0.586\pm0.061)\mu e\\
    &\quad+(0.150\pm0.041)e^2+(-0.198\pm0.074)\mu e^2]a_B.
\end{split}\end{equation}
Here $a_B$ is the binary separation. We adopt \citeauthor{hw99}'s convention for the mass ratio, letting $\mu$ denote the ratio $m_2/(m_1+m_2)$, where $m_1$ is the mass of the primary (the star the asteroid orbits), and $m_2$ is the mass of the secondary. Similarly, they give the minimum semi-major axis for stable P-type orbits as
\begin{equation}\label{eqn:ap}\begin{split}
    a_P &= [(1.60\pm0.04)+(5.10\pm0.05)e\\
    &\quad+(-2.22\pm0.11)e^2+(4.12\pm0.09)\mu\\
    &\quad+(-4.27\pm0.17)e\mu+(-5.09\pm0.11)\mu^2\\
    &\quad+(4.61\pm0.36)e^2\mu^2]a_B.
\end{split}\end{equation}
Here $\mu$ is taken to lie between $0$ and $0.5$, since the mass ratios $\mu$ and $1-\mu$ are equivalent up to labelling of the component stars.

More recently, extensive simulations were carried out by \citet{qsk+18}  for P-type orbits and \citet{qlk+20} for S-type orbits. Their results largely agree with those of \citet{hw99}, but cover the $(e,\mu)$ parameter space in greater detail, allowing for grid interpolation of the critical semi-major axes $a_S$ and $a_P$.

A different, more analytical, approach to determining the boundaries of the region of stable orbits was taken by \citet{szebehely80}, who made use of the fact that the quantity
\begin{equation}
    C_J=\frac{2GM_1}{r_1}+\frac{2GM_2}{r_2}+\omega^2r^2-v^2,
\end{equation}
called the Jacobi constant, is conserved in the circular restricted three-body problem. Here $M_1$ and $M_2$ are the masses of the two primary bodies, $\omega$ is their orbital frequency, and $r_1$ and $r_2$ are the distances between each and the third, small body, while $r$ is the small body's distance from the center of mass and $v$ is its velocity in the synodic frame (co-rotating with the binary). For a particular value of $C_J$, the surfaces corresponding to $v=0$ are called zero-velocity surfaces. They bound regions in space that a small body with that particular Jacobi constant cannot enter, since the square of its velocity must always be positive.

\citet{szebehely80} made use of this by calculating the Jacobi constant for initial conditions that, in the appropriate two-body approximation, would correspond to a circular orbit. S-type orbits were considered stable if the zero-velocity surface prevented the small body from escaping to infinity. Similarly, P-type orbits were considered stable if the zero-velocity surface prevented the small body from approaching arbitrarily closely to either of the primaries. This definition of stability, termed Hill-type stability \citep{szebehely78}, differs qualitatively from that of \citet{hw99}: on the one hand, it is global and not limited by a finite integration time or numerical precision; but on the other hand, it does not take into account all modes of instability. S-type orbits that are Hill stable may eventually result in a collision with the primary, while P-type orbits that are Hill stable may eventually escape. Nevertheless, the results obtained by \citet{szebehely80} for S-type orbits and by \citet{sm81} for P-type orbits are in broad agreement with the results of \citet{hw99}. 

We estimate the boundary of stability for P-type orbits using equation~(\ref{eqn:ap}) throughout the paper. Because the results of \citet{sm81} largely agree with those of \citet{hw99}, our results should not be sensitive to the particular criterion for stability adopted.


\clearpage
\bibliographystyle{aasjournal}
\bibliography{asteroid-belts}

\end{document}